%% file: main.tex
\documentclass[preprint,review,12pt]{elsarticle}

\usepackage{bm}
\usepackage{graphicx,url}

\usepackage{afterpage} 

\usepackage{floatflt}
\usepackage{latexsym}
\usepackage{fancyhdr}
\usepackage{longtable}
\usepackage{array}
\usepackage{epsfig}
\usepackage{epstopdf}
\usepackage{graphicx}
\usepackage{alltt}
\usepackage{verbatim}
\usepackage{amssymb,amsmath}
\usepackage{pifont}
\usepackage[usenames]{color}
\usepackage[rgb]{xcolor}
\usepackage{geometry}
\usepackage{enumitem}
\usepackage{cancel,soul,ulem}
\usepackage{caption}
\usepackage{subcaption}
\usepackage{afterpage}

\usepackage{comment}
\usepackage{amssymb}

\usepackage[utf8]{inputenc}

\include{definitions}

\journal{Computer Networks}

\begin{document}

\newcommand{\ntxtb}[1]{\textcolor{blue}{#1}}
\newcommand{\ntxtr}[1]{\textcolor{red}{#1}}

\begin{frontmatter}



\newtheorem{theorem}{Theorem}[section]
\newtheorem{conjecture}{Conjecture}[section]
\newtheorem{proposition}{Proposition}[section]
\newtheorem{assumption}{Assumption}[section]
\newcommand{\totallambda}{\lambda}
\renewcommand{\totallambda}{\lambda}

\newtheorem{lemma}{Lemma}[section]

\newtheorem{corollary}{Corollary}[section]
\newtheorem{remark}{Remark}[section]

\newtheorem{definition}{Definition}[section]
%
\newtheorem{example}{Example}[section]
\newlength{\labelexample}
\setlength{\labelexample}{1.8cm}
\renewenvironment{example}{\refstepcounter{example}\par\noindent{\gray\rule{\labelexample}{1pt}}\par\vspace*{-.2\baselineskip}\noindent\textsc{Example~\theexample:\xspace}}{\par\nobreak\vspace*{-.5\baselineskip}{\gray\hfill\rule{\labelexample}{1pt}}\par}

\title{Search and Placement in Tiered Cache Networks}


\author[label1,label2]{Guilherme Domingues}

\author[label1]{Edmundo de Souza e Silva}

\author[label1]{Rosa M. M. Leão}

\author[label1]{\\Daniel S. Menasché}

\address[label1]{Federal University of Rio de Janeiro, Brazil}

\author[label3]{Don Towsley}

\address[label2]{State University of Rio de Janeiro, Brazil}

\address[label3]{University of Massachusetts at Amherst, USA}

\input{abstract}


\end{frontmatter}

\newcommand{\totallambda}{\lambda}

\input{intro}

\input{related}

\input{system}
\input{model}
\input{optm}

\input{eval}

\input{discussion}

\input{concl}
\input{ack}

\appendix
\input{appendix}

\input{appendixstateful}











\section*{References}

\bibliography{main}

\end{document}

%% file: definitions.tex
\newcommand{\bA}{ {\bf A} }

\newcommand{\bQ}{ {\bf Q} }

\newcommand{\bupsilon}{ {\mbox{\boldmath $\upsilon$}} }

\newcommand{\bpi}{ {\mbox{\boldmath $\pi$}} }

\newcommand{\bdelta}{ {\mbox{\boldmath $\delta$}} }

\newcommand{\bnu}{ {\mbox{\boldmath $\nu$}} }

\newcommand{\bb}{ {\bf b} }
\newcommand{\bc}{ {\bf c} }

%% file: abstract.tex
\begin{abstract}
Content distribution networks have been extremely successful in today's Internet.  
 Despite their success, there are still a number of scalability and performance challenges that motivate
   clean slate solutions for content dissemination, such as content centric networking. 
   In this paper, we address 
   two of the fundamental 
   problems faced by any content dissemination system: content search and content placement.  
   We consider a multi-tiered, multi-domain hierarchical system wherein random walks are used to cope
   with the tradeoff between exploitation of known paths towards custodians versus opportunistic  exploration of replicas
   in a given neighborhood.     
    TTL-like mechanisms, referred to as reinforced counters, are used for content
   placement. We propose an analytical model to study the interplay between search and placement. The model
   yields closed form expressions for metrics of interest such as the average delay experienced by users
   and the load placed on custodians.  
  Then, leveraging the model solution we pose a joint placement-search optimization problem.
  We show that previously proposed strategies for optimal placement, such as the square-root allocation, follow
  as special cases of ours, and that a bang-bang   search policy  is optimal if  content allocation is given.
\end{abstract}

%% file: intro.tex
\section{Introduction}
\label{sec:intro}

Content distribution is in the  vogue.  
Nowadays, virtually everybody can create, distribute and download
content through the Internet.
It is estimated that video distribution  will alone account for up to 80\% of global traffic by 2017~\cite{ciscovni}. 
Despite the success of the current Internet infrastructure to support user demand, 
 scalability challenges 
 motivate clean slate approaches for content dissemination, such as information centric networking.

In information centric networks (ICNs), 
 {the focus is on  content, rather than on hosts}~\cite{ccn1,dona}.
 {Each content has an identification  and is associated to at least one custodian. 
Once a request for a content is generated it flows towards a custodian
through  routers equipped with caches, referred to as \textit{cache-routers}.
A request that finds the content stored in a cache-router
does not have to access the custodian. 
This 
alleviates the load at the custodians, reduces the delay to retrieve the content
and the overall traffic in the network.}
 {To achieve performance gains with respect to existing architectures,} in information centric networks 
cache-routers must efficiently and distributedly  determine 
how to route content requests and 
 {where to} place contents.

 ICN architectures, such as NDN~\cite{ccn1}, are promising solutions for the future Internet.  
 Still, it is unclear the scope at which the proposed solutions are feasible~\cite{realitycheck}.  
 Incrementally deployable solutions are likely to prevail~\cite{incrementally2}, 
and identifying the  simplest foundational 
  attributes of ICN architectures is essential while envisioning  their  Internet scale deployment.  
  
{The efficient management of  distributed storage resources in the network
coupled with  the routing  of requests for information retrieval
are  of fundamental importance \cite{icn-comm-survey-2013,kurosesurvey}}.
 {However,} the interplay between search and placement is still not well understood, 
 {and} there is a need to study search and placement problems under 
 a holistic perspective. 
 {In fact, an adequate framework within which   to assess
the overall performance gains that ICNs can provide is still missing~\cite{icn-comm-survey-2013}.
  
In this paper, we propose and study a simple  ICN architecture 
comprising of a {logical} hierarchy of cache-routers divided into 
 {tiers, where  each tier is  subdivided into one or more logical domains} (Figure~\ref{fig:domains2}). 
In-between domains, requests are routed from users towards custodians which are assumed to 
be placed at the top of the hierarchy.

 


 
\begin{figure}[h!]
\center
\includegraphics[scale=0.3]{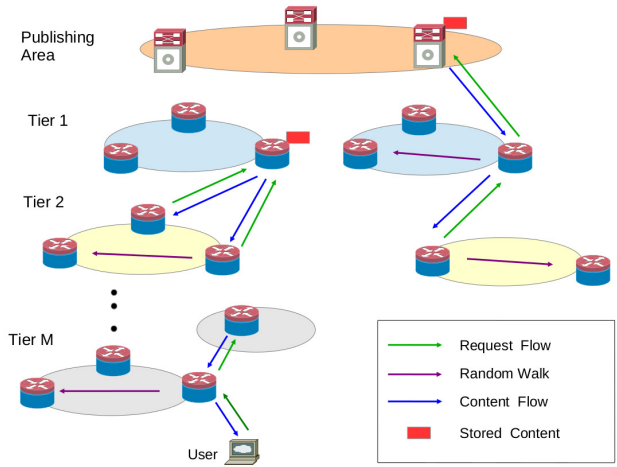}
\caption{System diagram}
\label{fig:domains2}
\end{figure} 
 
To route content requests from users to custodians, 
a random lookup search takes place 
in the vicinity of the logically connected cache-routers (horizontal arrows in Figure~\ref{fig:domains2}).
 {Cache-routers within a domain are assumed to form a logical \textit{clique}.
As such, a request that does not find the searched content in a cache-router is forwarded to
one of the remaining cache-routers in the same domain.}
The goal is to opportunistically explore the presence of content replicas in a given domain.    
If a copy is found in the domain 
 {within a reasonable time interval}, the content is  served. 
Otherwise, requests are routed from users towards custodians (vertical arrows in Figure~\ref{fig:domains2}).  
Custodians as well as the name resolution system (NRS) are supplied by third parties at 
 the publishing area, and we focus our attention on the infrastructure from users to publishing areas.

By using random walks to opportunistically explore the presence of content replicas closer to users,
we avoid  content routing tables and 
 tackle the scalability challenge posed in~\cite{challengesicn2015}.    An alternative would be to adopt  scoped-flooding~\cite{wangpro}.
 However, scoped-flooding is more complex than random walks and requires some level of synchronization between caches.
 In addition, random walks have been show to scale well in terms of overhead~\cite{ioannidis2008design}.
 

 
To  efficiently and distributedly place content in the cache network, we consider a flexible content placement 
mechanism inspired by TTL caches.  
At each cache, a counter is associated to each content stored there,  
 {which we refer to as  \textit{reinforced counter}} (RC). 
 Whenever the RC surpasses a given threshold, the corresponding content is stored.
  The RC  is decremented at a given established rate, until reaching zero, when the content is  evicted.



 Focusing on  two of the simplest possible mechanisms for search and placement, namely random walks and TTL-like caches,
our benefits are twofold.  From a practitioners  point of view, the proposed architecture 
is potentially deployable at the Internet scale~\cite{realitycheck}.  From the performance evaluation
 perspective, our architecture
 is amenable to analytical treatment.
Our  quantitative analysis provides closed-form expressions for
 different metrics of interest, such as the average delay experienced by users.

Given such an architecture, we pose the following  questions,
\begin{enumerate}

\item How long should the random-walk based search last at each domain so as to 
optimize the performance metrics of interest?

\item How should the reinforced counters be tuned so as to tradeoff content retrieval delay with server load at the custodian?

\item {What  parameters  have the greatest impact on the performance metrics
of the proposed ICN architecture?}

\end{enumerate}
 
To answer these questions, we introduce an analytical model that yields:
a) the  {expected delay to find a content} (average search time)
and; 
b) the rate at which requests  {have to be satisfied by  custodians}. 
While the expected delay is directly related to users quality of experience, 
the rate of accesses towards the custodian is associated with publishing costs.    
The model yields simple closed-form expressions for the metrics of interest. 


{Using the model, we study different tradeoffs involved in the setting of the parameter values.
In particular, we study the tradeoff between the time spent in opportunistic 
exploration around the vicinity of the  user  in order to find content and the custodian load.
}

In summary, our key contributions are the following:
    
\begin{description}

\item[{ICN architecture:}] we propose a simple ICN multi-tiered architecture based on random walks and TTL-like caches. 
Simplicity easies deployment and allows for analytical treatment, while capturing essential features of other ICN architectures
 such as the tension between opportunistic exploration of  replicas closer to users and exploitation of known paths towards custodians.   



\item[Analytical model:]
 {we introduce a simple analytical model of the proposed ICN architecture that can be helpful in the performance evaluation
of ICNs.
In particular, we consider the interplay between content placement and search.
Using the model we show that we can achieve performance gains using
a simple search strategy (random walks) and a logical hierarchical storage organization. }
Although our analysis is focused on the proposed architecture, we believe that the  insights  
obtained are more broadly applicable to other architectures as well,
such as scoped-flooding~\cite{wangpro}.


\item[{Parameter tuning}:]
we formulate an optimization problem that leverages  the closed-form expressions 
obtained with the proposed model to determine optimal search and placement parameters
 {under storage constraints}. 
We  show that previously proposed strategies for optimal placement, such as the square-root allocation, follow
  as special cases of our solution, and that a bang-bang   search policy  is optimal if  content allocation is given.

\item[Performance studies:] we 
 investigate how different parameters 
impact system performance under different assumptions regarding the relative rate at which
requests are issued and content is replaced in the cache-routers.  Our numerical investigations consider scenarios 
 in which the assumptions of the optimization problem  posed in this paper do not hold.
 

\end{description}

The remainder of this paper is organized as follows. 
After introducing background in Section~\ref{sec:related},
 we describe the system studied in this paper in Section~\ref{sec:system}.  
An analytic model of this system is presented in Section~\ref{sec:model}. 
The  joint placement and search optimization problem is posed and analyzed  in Section~\ref{sec:opt}
and numerical evaluations are presented in Section~\ref{sec:eval}.
Further discussions are presented in Section~\ref{sec:disc} and 
Section~\ref{sec:concl} concludes.

    

%% file: related.tex
\section{Background and Related Work}
\label{sec:related}

In this section we introduce the background used in this paper.  In Section~\ref{icnar}  we present previously 
proposed ICN architectures and in Section~\ref{chall} we indicate some of the challenges they pose.

\subsection{ICN Architectures} \label{icnar}

A survey comprising various architectures  considered for ICN can be
found  in~\cite{icn-comm-survey-2013}.   In what follows, we focus on five of  the 
 prominent 
architectures, namely DONA,
 PSIRP, Netinf, Multicache and NDN, which are most relevant to our work. 

DONA ~\cite{dona} consists  of a hierarchy of domains.  
Each domain includes a logical Resolution  Handler~(RH) that tracks  the 
contents published  in the domain and   in the descendant domains. 
Therefore, the logical RH placed in the highest level of the hierarchy 
is aware of all the content  published in  the entire  network. 
RHs provide a hierarchical name resolution service over the    routing
infra-structure. 
DONA supports  caching through the RH infrastructure.
When a RH aims at storing a content, it replaces the IP address of the requester
by its own IP address.
Then, the content will be delivered first to the RH before being 
forwarded to the end users, allowing the RH to cache the content   
within  the domain. 

PSIRP~\cite{psirp}, Netinf~\cite{netinf} and Multicache~\cite{multicache} 
handle  name resolution through a set of 
  Request   Nodes  (RNs)   organized   according  to   a  hierarchical
Distributed Hash  Table (DHT).  
Content  is sent to the  user through a set  of forward nodes
(FNs), under a  separate network. 
FNs can advertise cached information to RNs to enhance
the search efficiency and cache hit ratio. 
Nonetheless, as RNs cannot keep track of all replicas within
the network,  a key challenge consists of determining what is the relevant information to advertise.

The NDN~\cite{ccn1} architecture handles name resolution using content routing tables.
Users issue \textit{Interest} messages to request  a content. 
Messages are forwarded hop-by-hop by Content Routers~(CRs) until the content is found.
Messages leave a  trail of \textit{bread crumbs} and the content follows  the reverse 
path set by  the trail.  
As content flows to requesters,  the bread crumbs are removed. 
Published  content  is announced  through
routing protocols, with routing tables supporting  name aggregation (names are
hierarchical). 
To enhance  the discovery  of cached
contents, Rosensweig  et al.~\cite{elisha1} allow bread  crumbs not to
be  consumed on  the fly  when  content traverses  the network.  
This  allows trails for previously downloaded contents to be preserved.

\subsection{Challenges} \label{chall}

Some of the main challenges faced  by present ICN architectures 
are discussed in~\cite{challengesicn2015}. 
For Name Resolution Services~(NRS) lookup proposals, such as Dona, NetInf
and PSIRP, the challenge is to build a scalable resolution system which 
provides: 
(i) fast mapping of the name of the content to its locators;
(ii) fast update of the location of a content since locations
can change frequently;
(iii) an efficient scheme to incorporate copies of a content in the 
cache routers.

For  proposals based  on content routing  tables, such as  NDN,  the number  of
contents may  be around  $10^{15}$ to $10^{22}$.  
Routing table design becomes a challenge as its size is proportional to the number of contents in the system.
Route  announcements due to 
replica      updates,  and   link   failures,   pose additional 
challenges.

Simple hierarchical tiered topologies, wherein each domain comprises a single node, admit  closed-form expressions
for the expected time to access content~\cite{dabirmoghaddam2014understanding, towsley1}.  In this paper, we consider the case
where each domain comprises multiple nodes, which means that routing is non-trivial.    
To face the scalability  challenge related to content
 routing tables,~\cite{wangpro} proposes the use of flooding in each neighborhood, which
 simplifies design and reduces complexity.  
 In this paper, in contrast, we propose the use
 of random walks. Random walks  are as simple as flooding, and  lead to reduced congestion~\cite{gelenbe2010search, ioannidis2008design,lv2002search,gkantsidis2004random}  
 while still taking  advantage of spacial and temporal locality~\cite{traverso2013temporal, dabirmoghaddam2014understanding}.

For  proposals relying  on  DHTs there exist  many unsolved  security
vulnerabilities that are able to  disrupt the pre-defined operation of DHT 
nodes~\cite{securitydht}  and need to  be overcome. 
Note that in a network
composed  of   domains  where  providers   care  about  administrative
autonomy,    the    use   of    a    global    hash   table    becomes
unfeasible~\cite{mdht}.

%% file: system.tex
\section{System Architecture}
\label{sec:system}

{In this section} 
we describe the system architecture considered in this paper.  
{We begin with a brief overview}.

\subsection{Tiers and Domains}


The system consists of a set of cache-routers partitioned into several logical domains,
 which are organized into hierarchically arranged tiers  (Figure~\ref{fig:domains2}). 
%
%
%
%
Each domain consists of a set of routers or cache-routers that are responsible
for forwarding 
requests and caching copies of  contents. 
In what follows, we assume that all routers are equipped with caches, and use interchangeably the terms router and cache-router.

Users generate requests at the lowest level of the hierarchy.  These  requests flow across
{domains, following the tier hierarchy} towards the  publishing areas, at the top of the hierarchy.   
Figure~\ref{fig:domains2} displays routers forwarding requests towards a publishing area (green arrows). 
We consider  {$M$}
logical hierarchical tiers. Tier 1 is the top level tier and tier {$M$} 
 {is} the bottom level {constituted by routers that are ``closest''
to the users, i.e., which are the first to receive requests from users}. 
The publishing area knows how to forward a request to a publisher in case the content is not found 
in any of the tiers. 
{We adopt} a strategy that allows \textit{opportunistic} encounters between requests and replicas 
in a best-effort manner.

%
Each cache maintains a counter  (one per content), referred to as a {\it{reinforced-counter}}, to  
establish thresholds to  guide content placement at the  caches. 
Copies of popular contents may be cached in the  
routers.  
Whenever a request arrives to a domain, it generates a random walk to explore the domain, 
so as to  allow opportunistic  encounters with  the desired
content,  taking advantage of the temporal and geographical correlations  encountered by 
 popular requests~\cite{traverso2013temporal}.  
We rely on random walks in order to avoid the
 control overhead associated to  routing table updates
and the  drawbacks 
 of DHTs discussed in the previous section. 


\subsection{Random Walk Search} \label{sec:rws}

Random walks are one of the simplest search mechanisms  with the flexibility to 
account for  opportunistic encounters between  user requests   and 
replicas stored  within the domains (purple   arrows in Figure~\ref{fig:domains2}).  
Opportunistic encounters  satisfy requests without the need for them  to reach the publishing area. 
{A request that reaches the publisher area indicates that the corresponding content
was not found in any of the domains traversed by it.}


When a  request arrives  to a  domain, if the cache-router that receives the
request does not have the content, it starts a random walk search.
The random walk lasts for at most $T$  units of  time,  
only traversing  routers  in the domain.
A time-to-live (TTL) counter  is set  to limit the amount of search time  for a content 
\textit{within  a domain}.  
%
%
%
If the content has not been found by the time the TTL counter expires, the router that holds the 
request transfers it to the next  tier above it in the hierarchy.

%

As a request  is forwarded up the hierarchy, 
\textit{backward pointers} are deployed. 
These pointers are  named \textit{bread crumbs}.  
When content  is located in  the network,  
{two actions are performed: 
(a) the content is sent to the requester and
(b) the content is \textit{possibly} stored in the caches of the routers that first received the request
in each domain   (those that initiated the random walk at a domain).  
Action (b) is performed  if the reinforced counters associated with the given content at the considered cache-routers  
reach a pre-determined threshold.  Note that a cache-router may store contents that were found either in its own domain or in tiers above it. 
The publisher can perform action (a) by either directly sending the content to the
requester, or by following the reverse path of the request (blue  arrows in Figure~\ref{fig:domains2}), whichever is more efficient.} 
As  the content follows  the path of \textit{bread  crumbs}, the trail is  erased.

\subsection{Reinforced Counter Based Placement}
\label{sec:rcdesc}

We consider a  special class of content placement mechanisms, henceforth
referred to  as  reinforced counters  (RC), similar in spirit to TTL-caches~\cite{towsley1}.

Each published content  in the network is identified  by a unique hash
key \textsf{\textit{inf}}. 
All cache-routers  have a set of RCs,  one for  {each}  content.
 Reinforced counters are affected by exogenous requests and interdomain requests,
{\textit{but not by   endogenous requests inside a given domain}}, 
{that is, their values are not altered by the random walk search}.  


At any cache, the reinforced-counter associated to a given content 
is  increment by  one at every exogenous or interdomain request to that content, and  is
decremented  by one  at every tick of a   timer.  
The timer  ticks  at a rate of $\mu$ ticks per second. 

{Associated with each RC is a}
threshold $K$.
Whenever a request for   content
\textsf{\textit{inf}}  reaches  a   router,  either 
(i) an  already pre-allocated counter  for \textsf{\textit{inf}}
is  incremented {by one} in this router  
or  
(ii) a new RC is  allocated for \textsf{\textit{inf}}  and set to  one.  
If the value of the RC surpasses $K$, the content is stored 
{after \textsf{\textit{inf}} is found}.

RCs are  {decremented}
 over  time. 
Whenever the RC for  \textsf{\textit{inf}} is decremented from $K+1$ to $K$ 
the content is evicted {from the cache}.
The counter is deallocated {when it reaches zero}. 

Note that the RC dynamics of different contents are uncoupled {and the RC values are
independent of each other}.
{
Cache storage constraints are taken into account in the model 
by limiting the {\it average} number of replicas in each cache, which corresponds to soft constraints.    
Since hard constraints on the cache occupancy must be enforced, the RC threshold should be set in such a way that   
 the probability of a cache overflow is small~\cite{mostafa}. 
By limiting the fraction of time that each content is cached, reinforced counters take advantage of statistical 
multiplexing of contents in the system.
}

\subsection{Stateless and Stateful  Searches}

We consider two variants of random walk searches: stateless and stateful.  Under stateless searches,
requests  do not carry any information about
previously visited cache-routers. In other words, when a cache-router is visited, the
only information that is known is the content of the cache currently being visited.  
In a stateful search requests either a) remember the cache-routers that have been
visited or b) know ahead of time what routers to visit.  We assume that in stateful searches 
the searcher never revisits cache-routers.  The stateless and stateful searches  are studied in Sections~\ref{sec:fix} and~\ref{sec:dyn},
 respectively.  

%% file: model.tex
\section{Analytical Model} 
\label{sec:model}

{In this section we present an  analytical model to obtain performance metrics for the ICN architecture 
described in the previous section,  illustrated in Figure \ref{fig:domains2}.}
The model  takes into account 
 the {performance} impact of content  search 
through random walks and the cache management mechanism based on  reinforced counters. 

In particular, the model allows one to 
compute the
probability of  finding a content in a domain 
  and the mean time to find it.  
{Using the model we  }
show the benefit of a  hierarchical structure  and study the tradeoff between the 
storage requirements of the cache-routers 
{and the load that reaches the publishing area}. 

{When a request reaches a cache-router, the local cache is
searched and if the content is locally stored it  is immediately retrieved and sent to the user.  
If the content is not found, a random walk search starts in the domain. 
We assume that the random search takes $V$ time units 
per each cache-router visited where $V$ is an  exponentially distributed random variable  with 
rate $\gamma$.
}

{
Long search times can have an adverse effect on performance; hence, 
  a timer is set when the random walk starts to limit the search time.
The search can last for at most $T$ time units.
The search ends when the timer expires or the content is found, whichever occurs first.
As described in Section~\ref{sec:rws}, if the timer expires the user request is sent to the next cache-router 
  in the tier hierarchy,
and the process restarts.} Table~\ref{tab:notation} summarizes the notation used in the remainder of this paper.

\afterpage{
\begin{center}
\scriptsize
\begin{longtable}{l p{4.8in}}
\hline
\hline
Parameter & Description \\
\hline
$1/\gamma$ & average delay per-hop, i.e.,  average time for the random walk  to check \\
&  for a content at a cache and possibly forward the request \\ 
$\mathcal{C}(\hat{\Lambda}_c)$ & cost  incurred by custodian  
                  (measured in delay experienced by users) \\
$C$ & number of contents \\
$M$ & number of tiers \\
$N$ & number of caches in domain under consideration \\
$\lambda_{c,i}$ & arrival rate of exogenous and interdomain requests for  
      content $c$ at typical cache of domain $i$, $\lambda =\sum_{i=1}^M \sum_{c=1}^C \lambda_{c,i}$ \\  
$\Lambda_c$ &  exogenous arrival rate of requests for  $c$ at the network 
  (except otherwise noted,  exogenous requests are issued at tier $M$) \\  
\hline
\hline
Variable & Description \\
\hline
$L_c$ & number of replicas of content $c$ in tagged tier \\
$\pi_{c,i}$ & probability that content $c$ is stored at typical  cache at domain $i$ \\
$\alpha_{c,i}$ & $=1/\mu_{c,i}$ \\
\hline
\hline
Control variable & Description \\
\hline
$\mu_{c,i}$ & reinforced counter decrement rate for content $c$ at domain $i$ \\
$T_{c,i}$ & TTL for content $c$ at domain $i$, i.e.,  \\
& {maximum time to perform a random search for} content $c$ at domain $i$ \\
\hline
\hline
Metric & Description \\
\hline
$R_{c,i}(t)$   & probability of not finding content $c$ at tier $i$ by time $t$ \\ 
$D_{c,i}$ & delay incurred for finding content $c$ at tier $i$ \\
$D_{c}$ & delay incurred for finding content $c$ \\
$D$ & delay incurred for finding typical content \\
$\hat{\Lambda}_c$ & rate of requests   for content $c$  at the publisher \\
\hline
\caption{Table of notation. Note that subscripts are omitted  when clear from context.}
\label{tab:notation}
\end{longtable}
\end{center}

}

\subsection{ {Cache hit and insertion ratios} }

Consider a given tagged tier and cache-router in this tier. 
We assume that requests to content $c$ arrive to this cache-router
according to a Poisson process with rate  $\lambda_c$. 
{
$\lambda_c$ is also referred to as the  content popularity.
Recent work~\cite{mendonca2015} using three months of data collected from the largest VoD
provider in Brazil indicates that, during peak hours, the Poisson process is well suited 
to model the video request arrival process. 
%
In our numerical experiments, we rely on the Poisson assumption coupled with the Zipf distribution for popularities 
to characterize the workload.
}

{
}

{
We recall from Section \ref{sec:rcdesc}
that the reinforced counter associated to a given content $c$ is incremented at every request  
for $c$ and decremented at  constant rate $\mu_c$.  We assume that the counter is decremented at exponentially
 distributed times with mean $1/\mu_c$.  
Associated with each counter and content is a threshold   $K_c$
such that when the counter exceeds $K_c$, content $c$ must be stored
into cache.
}
{Let $\pi_c$ denote  the probability that the cache-router contains content $c$. 
Due to the assumption of Poisson arrivals and exponential decrement times, 
the dynamics of each reinforced counter is characterized by a birth-death process.   
}
Hence $\pi_c$, which is  the probability that  the reinforced counter has value greater than $K_c$, is given by
\begin{equation}
\label{eq:pic} 
\pi_c =  \left(\frac{\lambda_c}{\mu_c}\right)^{K_c+1}
\end{equation}
%
If $K_c=0$ we have $\pi_c = \lambda_c/\mu_c$, which we denote by $\rho_c$. 

Let $\beta_c$ denote the miss rate for content $c$. 
Then,
\begin{equation}
\label{eq:miss_r}
\beta_c= \lambda_c (1-\pi_c) 
\end{equation}

In~\ref{appcacheins}  we consider an additional  metric of interest, namely the cache insertion rate, which is
the rate at which content is inserted into cache.  Note that
the cache insertion rate is lower than the cache miss rate, as not all misses lead to content insertions.     
We show that  larger values of $K_c$ yield lower insertion rates, which translate into less overhead due to content churn.
{Despite the advantages of using $K_c>0$, without loss of generality,
and to facilitate the exposition,}   
in the remainder of this paper we assume $K_c=0$, except otherwise noted.


\subsection{Publisher Hit Probability}

We start by considering a single domain in a single tiered hierarchy, wherein $N$ cache-routers are logically fully connected,
{i.e., any cache-router can exchange messages with any other router in the same domain}. 
Our goal is to compute the probability 
{$R(t)$}
that a random walk does not find the requested
content by time $t$, $t>0$.  Note that  $R(T_c)$ equals the probability that 
  the request is forwarded to the custodian.  

{
We consider two slightly different models.
As in the previous section, both models assume that  requests for   a content arrive according to a Poisson process.
In what follows we describe the assumptions associated with each model, and comment on their
applicability.
}
In Sections~\ref{sec:fix} and~\ref{sec:dyn} the analysis of stateless and stateful searches focuses on  a tagged content~$c$. 
 
\subsubsection{Model 1: Stateless search}
\label{sec:fix}

Recall that a \textit{stateless search} is a search in which requests do not carry any information
about previously visited cache-routers. 
%
  We assume that  searches 
 are sufficiently fast so that the
probability that  content placement in a domain changes during the search is  negligible. 
This assumption is reasonable if the expected  time  it takes for the random walker to check for the presence of 
 content $c$ in a cache and 
to transit from a cache-router to another, $1/\gamma$,  is very small compared to the mean time between:
(a) two requests for $c$, $1/\lambda_c$, and;
(b) decrements of the reinforced counter for $c$, $1/\mu_c$.

{
When an inter-domain request for a given content $c$ arrives at a cache-router and a miss occurs, 
a random stateless search for $c$ starts. 
After  each visit to a cache-router, if the content is not found another cache-router is selected uniformly  at random
among  the remaining $N-1$ cache-routers.
Note that, because the search is stateless, nodes can be revisited  during the  search.
}

In Section~\ref{sec:rcdesc} we discussed the decoupling between RCs of different contents in a given cache.   
Next, we argue that RCs for different caches in a domain can also be treated independently. 
%
Recall that reinforced counters are not affected by   endogenous requests inside a given domain, 
so we restrict ourselves to the impact of 
inter-domain requests when studying cache occupancies.  
Due to symmetry, we  assume that the  rate of requests from outside of a domain for a given content at different cache-routers in a domain are 
 identical. 
 Due to the Poisson assumption, 
a request for content $c$ that arrives at a tagged cache-router 
sees the system in equilibrium (PASTA property).
Therefore, arrivals will find the content of interest at a given cache with probability $\pi_c$, independent of the state
 of the neighboring caches in that domain.  

Let $L_c$ be the random variable equal to the number of replicas of the content $c$ in the domain, 
excluding the router being visited.
We have:
\begin{equation}
\label{eq:Lrepl}
P(L_c=l) =  \binom {N-1} {l} \pi_c^{l}(1-\pi_c)^{N-1-l} .
\end{equation}

Let $J_c$ denote the number of hops traversed by the stateless request by time $t$.
Since the time between visits is assumed to be exponentially distributed,
\begin{eqnarray}
\label{eq:limit0}
R(t|J_c=j,L_c=l) &=&  (1 - \pi_c) (1-w_l)^{j}
\end{eqnarray}
where $w_l$ is the conditional probability that
the random walker selects one router with content $c$ from the remaining
$N-1$ routers in the domain 
when there are $l$ replicas of the content in the domain given that  the current router does not have the content.
Then, $w_l=l/(N-1)$.
 Note that $\pi_c$ depends on the placement policy and is defined partially by its parameter values.
\begin{proposition}
\label{prop:01}
The probability $R_c(t|L_c=l)$ is given by 
\begin{eqnarray}
\label{eq:rtm}
R_c(t|L_c=l) &=& (1 - \pi_c)  e^{-\gamma \omega_l t}
\end{eqnarray}
\end{proposition}
\textit{Proof: }
From \eqref{eq:limit0}  we have:
\begin{eqnarray}
R_c(t|L_c=l) &=& 
   (1 - \pi_c) \sum_{n=0}^{\infty} \frac{(\gamma t)^n}{n!}(1-\omega_l)^n e^{-\gamma t} \nonumber \\
     &=& \frac{1 - \pi_c} {e^{\gamma t \omega_l}}
       \sum_{n=0}^{\infty} \frac{(\gamma t (1-\omega_l))^n}{n!}e^{- \gamma t (1-\omega_l)} \nonumber \\
     &=& (1 - \pi_c)e^{-\gamma \omega_l t} 
\end{eqnarray}
$\square$

\begin{proposition}[Stateless search]
\label{prop:Rt}
The probability ${R}_c(t)$ that a walker does not find a requested tagged content
{in a domain}
by time $t$ is given by: 
\begin{eqnarray}
\label{eq:rt} 
R_c(t) &=&
  { \left( { e^{-\gamma t / (N-1)} \pi_c + (1-\pi_c)  } \right) }^{(N-1)} (1-\pi_c)
\end{eqnarray}
\end{proposition}
\textit{Proof:}
Unconditioning \eqref{eq:rtm} on $L_c$,  yields 
\begin{eqnarray}
R(t) &=& \sum_{l=0}^{N-1} R(t|L_c=l) \binom {N-1} {l} \pi_c^{l} (1-\pi_c)^{(N-1-l)} \nonumber \\
&=&  (1-\pi_c) \sum_{l=0}^{N-1} e^{-\gamma \omega_l t} \binom {N-1} {l} \pi_c^{l}(1-\pi_c)^{(N-1-l)} \nonumber \\
&=&  (1-\pi_c) \sum_{l=0}^{N-1} \binom {N-1} {l} \left( {e^{-\gamma t/(N-1)} \pi_c }\right)^{l} (1-\pi_c)^{N-1-l} \nonumber \\
&=&  (1-\pi_c)  \left( { \pi_c e^{-\gamma t / (N-1) } + (1-\pi_c)  } \right)^{(N-1)} 
\end{eqnarray}
$\square$

According to \eqref{eq:rt}, $R_c(\infty) = (1-\pi_c)^{N}$.  
As  $t$ increases, the probability that the walker does not find  
content $c$ approaches the probability that all $N$  caches within the domain do not hold the  content. 

\label{sec:model1}

\subsubsection{Model 2: Stateful search} 
\label{sec:dyn}


In this section, we consider 
  \textit{stateful searches}  wherein requests  remember  the  
cache-routers that have been  visited, i.e., after the search is initiated, the searcher
chooses the next router to visit uniformly at random, from those that have not yet
been visited before.  
Alternatively,   
 requests  
know ahead of time what routers to visit. 
 This latter approach is discussed in~\ref{sec:appstateful}.
 

Under a stateful search, the searcher never revisits cache-routers. 
This is possible because cache-routers are logically fully-connected.  
As in the stateless model, we assume that arrivals of inter-domain requests for content $c$   
at  cache-routers are characterized by Poisson processes.
Therefore, the random searches for $c$ that are initiated at a tagged router $i$ 
are  characterized by a Poisson process modulated by the RC  of router $i$, 
whose dynamics is governed  by  a birth-death Markovian process.
It is shown in~\cite{pasta-ext1992} that the PASTA property holds for Poisson processes modulated by independent Markovian 
processess.  Therefore,   
a search that starts at router $i$ and  arrives at  router 
$k \neq i$
sees the RC  at $k$  in equilibrium, i.e.,  the request issued at router $i$ finds the desired content at cache  $k$
 with probability $\pi_c$.  Conditioning on  $J_c=j$ hops being  traversed by time $t$, the probability 
 that content $c$ is not found  is given by 
%
\begin{eqnarray}
\label{eq:Rt-j-2}
\tilde{R}_c(t | J_c=j) &=&  (1-\pi_c)^{j+1}
\end{eqnarray}
It remains to remove the conditioning  on $J_c$. 

We  assume, as in the stateless  model, that the search takes an exponentially distributed random delay at each hop,
 independent of the system state.
%
%
\begin{proposition}[Stateful search]
\label{prop:model2}
The probability $\tilde{R}(t)$ that a tagged content is not found by a stateful search by time $t$
 is given by
\begin{equation}
\tilde{R}_c(t) =  (1-\pi_c)( e^{-\gamma \pi_c t} + g(N)) 
\end{equation}
where 
\begin{equation}
g(N) = (1-\pi_c)^{N-1}\sum_{n=N}^{\infty} \frac{(\gamma t)^n} {n!} e^{-\gamma t} \left(1-(1-\pi_c)^{n+1-N}\right)
\end{equation}
\end{proposition}

\textit{Proof:}
{
The proof is similar to that  of Proposition~\ref{prop:01}.
The time between cache visits is an exponential random variable with rate $\gamma$.  
It follows from \eqref{eq:Rt-j-2} that  
\begin{eqnarray}
\tilde{R}_c(t) &=& \sum_{n=0}^{N-1} \tilde{R}_c(t|J=n) \frac{(\gamma t)^n}{n!} e^{-\gamma t} +  \tilde{R}_c(t|J=N-1) \sum_{n=N}^{\infty} \frac{(\gamma t)^n}{n!} e^{-\gamma t} 
\\
  &=& (1-\pi_c)\left( \sum_{n=0}^{N-1} \frac{(\gamma t)^n}{n!} e^{-\gamma t} (1-\pi_c)^n  +  (1-\pi_c)^{N-1} \sum_{n=N}^{\infty} \frac{(\gamma t)^n}{n!} e^{-\gamma t} \right) \nonumber \\
   &=& (1-\pi_c)\left( \sum_{n=0}^{\infty} \frac{((1-\pi_c) \gamma t)^n }{n!} \frac{e^{-\gamma(1-\pi_c)t}}
       {e^{\gamma \pi_c t}} + g(N) \right) \nonumber \\
   &=& (1-\pi_c) \left( e^{-\gamma \pi_c t} + g(N) \right)
\end{eqnarray}
$\square$
%
}

{
For large values of $N$, it follows from Proposition~\ref{prop:model2} that 
\begin{equation}
\tilde{R}_c(t) \approx  (1-\pi_c) e^{-\gamma \pi_c t} \label{eq:rtm1} 
 \end{equation}
 The validity of the large $N$ assumption 
can be checked by using the Normal distribution approximation for the Poisson distribution.
 For instance, the sum $\sum_{n=N}^{\infty} \frac{(\gamma t)^n} {n!} e^{-\gamma t}$ that appears in the expression of $g(N)$
 is well approximated by 
 the complementary cumulative distribution of the  Normal distribution,  
 $1-\Phi\left(\frac{N-\gamma t}{\sqrt{\gamma t}}\right)$,   for values of $N > \gamma  t + 4 \sqrt{\gamma t}$, where 
 $\Phi(x)$ is the cumulative distribution function of the standard Normal distribution. 
}

According to ~\eqref{eq:rtm1}, $\tilde{R}_c(\infty) = 0$.  
As the random walk progresses, 
contents are dynamically inserted and evicted from  the
caches and the walker eventually finds the desired content.


\subsubsection{Multi-tier {Networks}}

{
In the previous sections we considered a single tiered network.
In what follows we extend these results to the multi-tier case.
In Section \ref{sec:eval} we discuss the potential performance benefits of a multi-tiered architecture.
} 
 
{Refer to Figure \ref{fig:domains2} and}
let $M$ denote the number of tiers.  
Let $\hat{\Lambda}_c$ denote the publisher load accounting for the requests filtered at the $M$ tiers.
Let $R_{c,i}(T_{c,i})$ denote the probability that a search that reaches domain $i$ 
fails to find  content $c$ at that domain.   
The load for content $c$ that arrives at the publishing area is given by:
\begin{eqnarray}
\hat{\Lambda}_c =\Lambda_c \prod_{i=1}^{M} R_{c,i}(T_{c,i})  
\end{eqnarray}
where $\prod_{i=1}^{M} R_{c,i}(T_{c,i})$
is the probability that a request arrives at the publishing area
and $\Lambda_c$ is the  load generated by the
users for content $c$ which are all placed at tier $M$. 
Note that replacing $R_{c,i}(T_{c,i})$ 
by $\tilde{R}_{c,i}(T_{c,i})$ 
corresponds to 
{using the stateful model in place of the stateless  one.}

\subsection{Average Delay}

Let ${D}_{c,i}$ be a random variable denoting the delay experienced by requests for 
content $c$ at domain $i$.  
Recall that  $T_{c,i}$ is the maximum time a walker spends searching for content $c$ in 
domain $i$. 
In what follows, we make the dependence of $D_{c,i}$ on $T_{c,i}$ explicit.
It follows from~\cite{transol2000} that
%
\begin{eqnarray}
E[{D}_{c,i}(T_{c,i})] = \int_{0}^{T_{c,i}} R_{c,i}(t)dt \label{eqlt}
\end{eqnarray}
{Under the stateless model, }
$E[{D}_{c,i}(T_{c,i})]$ does not admit a simple closed form solution and must be obtained
through numerical integration of~\eqref{eq:rt}.    
{On the other hand, when the stateful model is employed, we obtain, after replacing~\eqref{eq:rtm1} into~\eqref{eqlt},}
\begin{eqnarray}
E[{D}_{c,i}(T_{c,i})] &=& \int_{0}^{T_{c,i}} (1-\pi_{c,i})   e^{ -\gamma \pi_{c,i} t}dt  \\
&=&  (1-\pi_{c,i})\frac{1-e^{-\gamma \pi_{c,i} T_{c,i}}}{\pi_{c,i} \gamma}. \label{eqlt1}
\end{eqnarray}

Let $D_c$ denote the  delay to find  content $c$, 
including the time required for the publishing area to serve the request if needed. 
Then, $E[D_c]$ is given by:
\begin{eqnarray}
\label{eq:edd}
E[D_c] = \left(\sum_{i=1}^{M} E[D_{c,i}(T_{c,i})]\prod_{j=i+1}^{M} R_{c,j}(T_{c,j}) \right) +
\mathcal{C}(\hat{\Lambda}_{c}) \prod_{j=1}^{M} R_{c,j}(T_{c,j}),
\end{eqnarray}
where $\mathcal{C}(\hat{\Lambda}_{c})$ is the mean cost (measured in time units) 
to retrieve a content at the 
publishing area as a function of the load $\hat{\Lambda}_{c}$. 
Recall that tier 1 (resp., tier $M$) is the closest to the custodians (resp., users).
Therefore, $\prod_{j=i+1}^{M} R_{c,j}(T_{c,j})$ corresponds to
the fraction of requests to content $c$ that reach tier $i$, for $i=1, \ldots, M-1$. 


%% file: optm.tex
\section{{Parameter Tuning}}
\label{sec:opt}

In this  section we consider the problem of minimizing average delay
under  average storage  constraints.  
To  this aim,  we use  
the stateful model { that was}
introduced  in the previous  section.
While
{in Section \ref{sec:model} }
the analysis targeted a single
tagged  content, in  this section  we account for the 
{limited space available in the caches and for contents that compete for cache space.}
%
 
To simplify presentation, we consider a single tier ($M=1$).  
We also assume  that the  delays experienced 
by  requests at the custodian are  given and fixed, equal to $\mathcal{C}$.
 
Let $D_c$ denote  the delay experienced by a requester of content $c$.    
$E[D_c]$ is obtained by substituting  \eqref{eq:rtm1} into \eqref{eq:edd},
 \begin{equation} \label{eq:dc}
 E[D_c] = (1-\pi_c)\left( \frac{1-e^{-\gamma \pi_c T_c }}{\pi_c \gamma}  + \mathcal{C} e^{-\gamma \pi_c T_c}  \right) 
 \end{equation}
 and
 \begin{equation} \label{eq:d}
 E[D] = \sum_{c=1}^C \frac{\lambda_c}{\totallambda} E[D_c]
 \end{equation}
 
Let $\alpha_c = 1/\mu_c$,  
$\bm{\alpha} = (\alpha_1, \alpha_2, \ldots, \alpha_C)$ and 
$\bm{T} = (T_1, T_2, \ldots, T_C)$.   
In light of~\eqref{eq:pic} and ~\eqref{eq:dc}-\eqref{eq:d},  
{we}
pose the following 
{joint placement and search}
optimization problem: 
\begin{eqnarray}
\label{eq:joint-opt}
\min_{(\bm{\alpha},\bm{T})} && E[D]= \sum_{c=1}^C \frac{\lambda_c}{\totallambda} (1-\lambda_c \alpha_c) \left(\frac{1-e^{-\gamma \lambda_c \alpha_c T_c }}{\lambda_c \alpha_c  \gamma}  + \mathcal{C} e^{-\gamma \lambda_c \alpha_c  T_c}  \right) \nonumber \\ 
s.t. && \sum_{c=1}^{C}\lambda_c \alpha_c = B
\end{eqnarray}
Note that  we impose a constraint on the  expected buffer size, i.e.,
the number of expected items in the cache cannot exceed the buffer size $B$. 
Similar constraint has been considered, for instance, in~\cite{melazzi2014general}.
Moreover, recent work~\cite{mostafa} shows that, for TTL caches,  we can size the buffer as $B(1 + \epsilon)$,  
where $B$ (resp., $\epsilon$) grows in a sublinear manner (resp., shrinks to zero) with respect to $C$, and 
 content will  not need to be evicted from the cache before their timers expire, with high probability. 


The reinforced counter vector  $\bm{\alpha}$  impacts content 
placement, 
while the random walk vector $\bm{T}$ impacts content search.  
By jointly optimizing for placement and search parameters, under
storage constraints, we obtain insights about the interplay between these two fundamental mechanisms.

{
In what follows, we do not solve the joint optimization problem directly.
Instead, to simplify the solution, we solve  two  problems independently:
first,  we consider the optimal placement given a search strategy, and then the optimal search 
given a pre-determined placement.
In our case studies we discuss the impact of these simplifications. 
}

\subsection{Optimal Placement Given Search Strategy}
\label{sec:optplgrout}

We first address the optimal placement problem, that is we determine 
how the buffer space at the cache-routers should be statistically divided
among the contents to optimize the overall performance.

%
  
\subsubsection{Special Case: $T$ large}

We begin by considering  large time to live values.  In the limit when $T_c=\infty$, the  time spent  locally searching
  for a content
is unbounded.
Under this assumption, the optimization problem stated in \eqref{eq:joint-opt} reduces to 
\begin{eqnarray}
 \min_{\bm{\pi}} 
    &&  \sum_{c=1}^C \frac{\lambda_c}{\totallambda} ( 1-\pi_c)
       \left(\frac{1} {\pi_c  \gamma} \right) \nonumber \\
  s.t. && \sum_{c=1}^{C}\pi_c = B \label{eq:constr1}
 \end{eqnarray}


We construct the  Lagrange function,
\begin{equation}
\mathcal{L}(\bm{\pi},\beta) =
   \sum_{c=1}^C \frac{\lambda_c}{\totallambda} \frac{( 1-\pi_c)}{\pi_c \gamma} 
   + \beta \left( \sum_{c=1}^C \pi_c - B \right)
\end{equation}
where $\beta$ is a Lagrange multiplier. 
Setting the derivative of the Lagrangian with respect to $\pi_c$ equal to zero and
using~\eqref{eq:constr1} yields, 
\begin{equation}
\beta=\frac{\left(\sum_{c=1}^C \sqrt{\lambda_c}\right)^2 }{ \gamma \lambda B^2}.
\end{equation} 
Therefore, 
\begin{equation}
\pi_c = B {\frac{\sqrt{\lambda_c}}{\left(\sum_{c=1}^C \sqrt{\lambda_c}\right)}}, c=1, \ldots, C. \label{squareroot}
\end{equation}

When $B=1$, the optimal policy~\eqref{squareroot}
is the square-root allocation proposed by Cohen and Shenker~\cite{cohen2002replication} 
in the context of peer-to-peer systems.  
It is interesting that we obtain  a similar result for the ICN system under study. 
This is because in both cases the optimization problem  can be reformulated
as  to   minimize 
$ \sum_{c=1}^C (\lambda_c/\lambda)/\pi_c$ under the constraint that $\sum_{c=1}^C \pi_c =B$.  In~\cite{cohen2002replication}
 the term $1/\pi_c$ is the mean time to find  content $c$, which is the average of a geometric random variable with 
 probability of success $\pi_c$.  In the ICN system under study, the term $1/\pi_c$   
 follows from  expression~\eqref{eq:dc}. 

\subsubsection{Special Case: $ T=0$}
\label{sec:optplgrout}

Next, we consider the case $T=0$. 
When a request for $c$ arrives at a cache-router and does not find the content, the request is automatically 
sent to the next
level in the hierarchy of tiers.
Then, the optimization problem reduces to 
\begin{eqnarray}
\min_{\bm{\pi}} && E[D]= \sum_{c=1}^C \frac{\lambda_c}{\totallambda} ( 1-\pi_c) \mathcal{C}\\ 
s.t. && \sum_{c=1}^{C}\pi_c = B 
\end{eqnarray}
 
In this case, the optimal solution consists of ordering contents based on $\lambda_c$
and storing the $B$ most popular ones in the cache, i.e., $\pi_c=1$ for $c=1, \ldots, B$ and $\pi_c=0$ otherwise.
Note that this  rule was shown to be optimal
by Liu, Nain, Niclausse  and Towsley~\cite{liu1997static} in the context of Web servers.

\subsubsection{Special Case: $\gamma T$ small} 
 \label{sec:gammatsmall}
 
For $\gamma T << 1$, we have $e^{-\gamma \pi_c T} \approx 1-\gamma \pi_c T$.  The optimization problem is given by 
 \begin{eqnarray}
 \min_{\bm{\pi}} && E[D]= \sum_{c=1}^C \frac{\lambda_c}{\totallambda} ( 1-\pi_c)\left(T+(1-\gamma \pi_c T) \mathcal{C} \right) \\ 
  s.t. && \sum_{c=1}^{C}\pi_c = B \\ 
  && 0 \leq \pi_c \leq 1
 \end{eqnarray}
 Note that the objective function can be rewritten as
\begin{equation} 
 E[D]= \sum_{c=1}^C  \frac{\lambda_c}{\lambda} \left( \mathcal{C} T \gamma \pi_c^2 + \left(-(\mathcal{C}+T)- \mathcal{C} T \gamma \right) \pi_c + (\mathcal{C}+T)\right)
\end{equation}
This is a special separable 
convex quadratic program, known as the economic
dispatch problem~\cite{bay2010analytic} or  
continuous quadratic knapsack~\cite{gallo1980quadratic}.  
 It can be solved in linear time
using  techniques presented in~\cite{bayon2013exact}.  
 Alternatively, in~\ref{appsmallgammat} we present the dual of the problem above, 
 which naturally yields a simple interactive gradient descent solution algorithm.

\subsection{Optimal Search Given Placement} 
\label{sec:optroutgpl}
 
In this section we address the optimal search problem, that is the choice of the $T_c$'s,
when placement is given (the $\pi_c$'s have been determined).
%
%
Then, the problem reduces to
\begin{eqnarray} \label{minfixpl}
\min_{\bm{T}}  && \sum_{c=1}^C \frac{\lambda_c}{\lambda} \left(1-\pi_c\right)\left(\frac{1-e^{- \gamma \pi_c T_c }}{\gamma \pi_c}  + \mathcal{C} e^{-\gamma \pi_c T_c} \right)\\
s.t. && T_c \ge 0, c=1, \ldots, C
\end{eqnarray}

For each content $c$ the function to be minimized is $f(T)$,
\begin{equation}
f(T)=\frac{1}{\gamma \pi_c} \left(1-e^{- \gamma \pi_c T}\right) + \mathcal{C} e^{- \pi_c \gamma T} \label{myobj}
\end{equation}
and
\begin{equation}
\frac{df(T)}{dT}= e^{- \gamma \pi_c T} - \gamma \pi_c \mathcal{C} e^{- \gamma \pi_c T}
\end{equation}

For a given content $c$, a random walk search should be issued with $T=\infty$ whenever 
${df(T)}/{dT} < 0$, i.e., if $1 - \gamma \pi_c \mathcal{C} < 0$.
Otherwise,  the request for content $c$ should be sent directly to the publishing area; 

\begin{equation}
\label{eq:Tc}
T_c =\left\{ \begin{array}{ll}
\infty, &  \pi_c > {1}/({\mathcal{C} \gamma}) \\
0, & \textrm{otherwise} 
\end{array} \right. 
\end{equation}
 
\textbf{Remarks}:
Although we do not solve the joint placement and search optimization problem, the special cases
considered above provide some guidance for system tuning.
The studies we conduct in the following section provide evidence of the usefulness of our model solutions.
In addition, we may try different approximation approaches to solve the combined placement and search problem.
For instance, one such approach is to first optimize for the $\pi_c$'s assuming $T$ is large
 and set $T_c = \infty$ for all contents  that satisfy 
  $\pi_c > {1}/({\mathcal{C} \gamma})$ (see \eqref{eq:Tc}). 
Then, set $T_c=0$ for the contents for which   $\pi_c \leq  {1}/({\mathcal{C} \gamma})$, 
and recompute  $\pi_c$ for such contents using the solution presented in Section~\ref{sec:optplgrout} so as to fill
the available buffer space. 
%
The performance of this and other heuristics is subject for future research.

%% file: eval.tex
\section{Evaluation}
\label{sec:eval}


In this section we report numerical results obtained using  the proposed model.  
Our goals are
a) to show tradeoffs involved in the choice of the \textit{time to live} (TTL) parameter,
b) to illustrate the interplay between content
search and placement, and
c) to numerically solve the optimization problems posed in this paper,  giving insights about the solutions. 
In Sections~\ref{sec:tradeoffttl}
and
\ref{sec:3-contents} we consider the stateless model, and in Section~\ref{sec:loadagg} 
 we consider the stateful one.

\subsection{Tradeoff in The Choice of TTL: Single Content Scenario}
\label{sec:tradeoffttl}

\begin{figure}[h!]
\center
\includegraphics[width=0.8\textwidth]{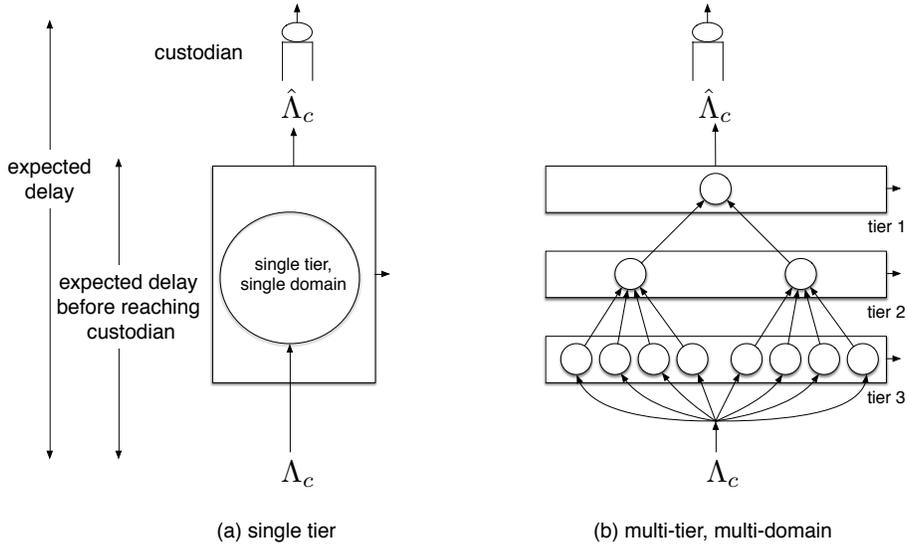}
\caption{Illustrative topology }
\label{fig:3tiers} 
\end{figure}

In this section we consider  a single content that is to be served in the three-tiered topology shown in Figure~\ref{fig:3tiers}(b).
Let $\Lambda_c=1$.   
We assume that  the number of replicas of the content
remains fixed while the walker traverses each domain (Section~\ref{sec:fix}).   
In addition, we assume that $\pi$ and $T$ are equal at the three considered domains (this assumption will be removed in the other considered scenarios). 
As requests are filtered towards the custodian, the rate of requests  decreases when moving from tier 3  to tier 1. 
The rate at which reinforced counters are decremented  also decreases, in order to keep $\pi$ constant.  

\begin{figure}[h!]
\center
\includegraphics[scale=0.30]{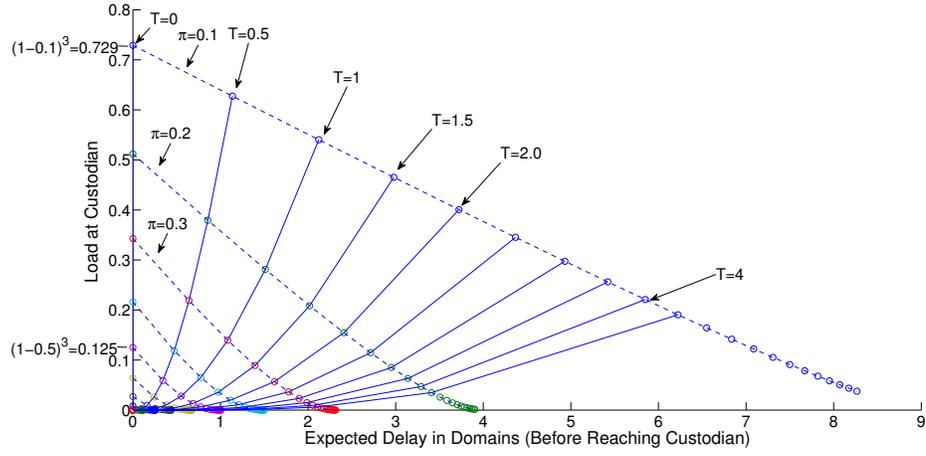} \\
(a) \\
\includegraphics[scale=0.30]{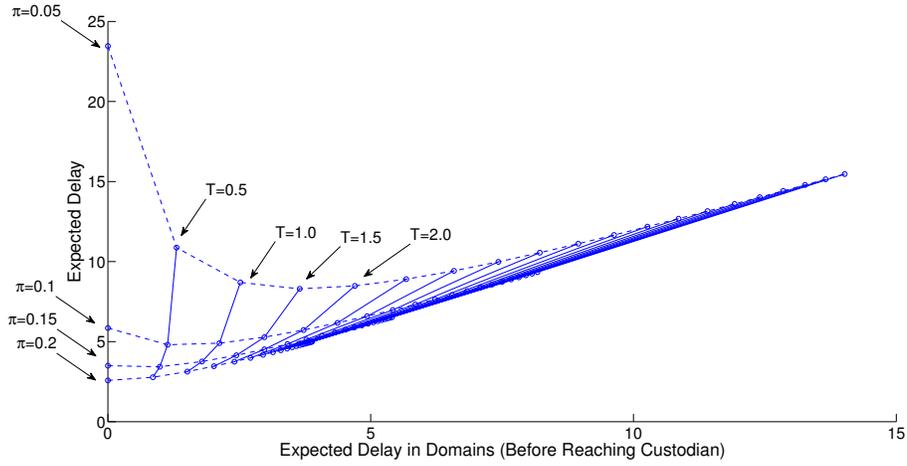} \\
(b)
\caption{Scatter plot indicating the tradeoff in the choice of TTL $T$: (a) larger values of $T$ reduce load in custodian at cost of increased expected delay in domain; (b) expected delay as a function of expected delay in domains, assuming cost at custodian $\mathcal{C}(\hat{\Lambda}_c)=1/(0.9-\hat{\Lambda}_c)$. }  \label{sec:resultsttl}
\end{figure}

Figure~\ref{sec:resultsttl}(a)  shows the 
 the expected delay to  
reach the custodian  and the custodian 
load     
for different values of  $\pi$ and $T$.  
For a given value of $\pi$, the dotted lines indicate that as $T$ increases the load at the custodian 
decreases and the expected delays in the domains increases.  
In contrast, for a given value of $T$, as $\pi$ increases, content becomes more available, which causes 
a decrease in the load at the custodian and in the expected delay.  

Next, our goal is to  evaluate the expected  delay. To this aim, we use an M/M/1 queue to model the 
delay at the  custodian.  
We let the custodian cost be given by $\mathcal{C}(\hat{\Lambda}_{c})=1/(0.9-\hat{\Lambda}_c)$, which corresponds
to the delay of an M/M/1 queue with service capacity of 0.9.

Figure~\ref{sec:resultsttl}(b) shows how the expected delay (obtained with equation~\eqref{eq:edd}) varies
as a function of  $\pi$ and $T$. 
For $\pi=0.05$ and $\pi=0.1$, as $T$ increases, the expected delay $E[D_c]$ first decreases
and then increases.  The initial decrease occurs due to a decrease in the custodian load.  
Nonetheless, as $T$ further increases the gains due to  decreased load at the custodian are dominated 
by the increased expected delay before reaching the custodian.
The optimal value of $T$ is approximately 1.5 and 0.5 for $\pi$ equal to 0.05 and 0.1, respectively.


\subsection{Benefits  of Load Aggregation}
\label{sec:loadagg}

\begin{figure}[h!]
\center
\includegraphics[scale=0.5]{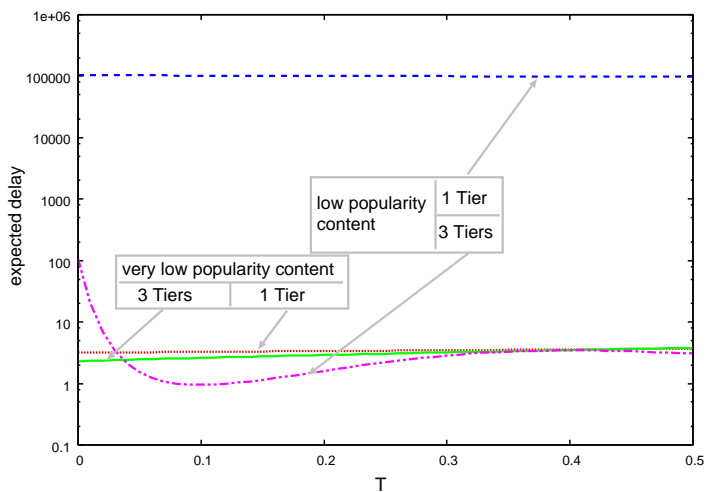}
\vspace{-6cm}
\caption{Expected Delay: very low and low popularity contents}
\label{fig:delay-lambda-001-01}
\end{figure}

\begin{figure}[h!]
\center
\includegraphics[scale=0.5]{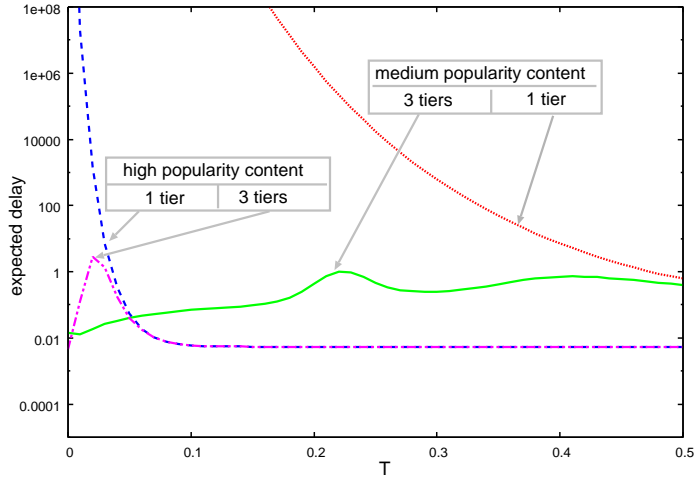}
\vspace{-6cm}
\caption{Expected Delay: medium and high popularity contents}
\label{fig:delay-lambda-05-09}
\end{figure}

While in the previous section we studied the dynamics of  a single content, now we 
consider four  content popularities: very low, low, medium and high. 
In Figures~\ref{fig:delay-lambda-001-01} and~\ref{fig:delay-lambda-05-09} 
we plot  expected delay both for the one-tiered architecture (Figure~\ref{fig:3tiers}(a)) and  
the three-tiered architecture (Figure~\ref{fig:3tiers}(b)).
The request arrival rate for each type of 
content was obtained from real data 
collected from a major Brazilian broadband service provider~\cite{mendonca2015}. 
The content request rates are $\lambda_1=0.8$, $\lambda_2=0.5$, $\lambda_3=0.1$ 
and  $\lambda_4=0.01$ req/sec.
The Request Counter (RC) of each content is decremented at  constant rate $\mu=1$ in  the three tiers. 
The value of $\pi$ varies for each content in each tier due to the fact that content requests are filtered out
as they travel towards the custodian.
In addition, we assume that $T$ is equal in the three  domains,
the number of replicas of each content
remains fixed while the walker traverses each domain (Section~\ref{sec:fix}), and     
the mean time to retrieve a content from the publishing area
   exponentially  increases with respect to  the  amount of requests 
hitting the publishing area, $\mathcal{C}(\hat{\Lambda}_c)=e^{\hat{\Lambda}_c}$.  



Figures~\ref{fig:delay-lambda-001-01} and~\ref{fig:delay-lambda-05-09} 
show the benefits of  load aggregation that
occurs in the three-tiered architecture:  requests that are not 
satisfied in tier three 
 are 
aggregated in the second and third tiers. Aggregation increases the probability to 
find the content in these tiers. 
%
%
We observe that contents with  low and medium popularities benefit the most from load  aggregation.
Note that the expected delay decreases by 
several orders of magnitude for low popularity contents  when we consider a three-tiered architecture.
For very low and high popularity contents, a significant reduction is not observed.
For highly popular contents, the probability to store the content in at least one of the tiers 
is high in both architectures,  
and only a small fraction of the requests is served by the publishing area.
For very low popularity contents, the opposite occurs: the majority of requests 
are served by the publishing area, as 
the probability that content is stored  in one  of the  tiers is very low.



Figures~\ref{fig:delay-lambda-001-01} and~\ref{fig:delay-lambda-05-09} show that the three-tiered architecture yields 
 lower delays, 
for all content popularities.
Next, we consider the optimal TTL choice in the three-tiered topology. 
For very low popularity contents, the best choice is  $T=0$ as   
the majority of requests must be served by the publishing area.
For high popularity contents, the best choice is also $T=0$ because
the probability to find the content in the first router of the domain is very high.
On the other hand, for low popularity contents, Figure~\ref{fig:delay-lambda-001-01}  shows that the  mean 
  delay is minimized  when 
$T \approx 0.1$.


\subsection{Validation of the Optimal Solution}
\label{sec:3-contents}

In this example our goal is to obtain the values of $\pi_c$ and $T_c$,
$c=1,2,3$, that minimize expected delay.
We consider three contents with high, medium 
and low popularity sharing a memory that can  store, on average,
one replica of  content, $B=1$. 
The publisher cost is $\mathcal{C}=10$, the random search time
is $1/\gamma=40$ ms and the content request rates are $\lambda_1=0.8$, $\lambda_2=0.1$
and $\lambda_3=0.002$ req/sec.  
As in the previous section, 
content popularities were inspired by data  
collected from a major Brazilian broadband service provider~\cite{mendonca2015}.

\begin{figure}[h!]
\center
\includegraphics[scale=0.5]{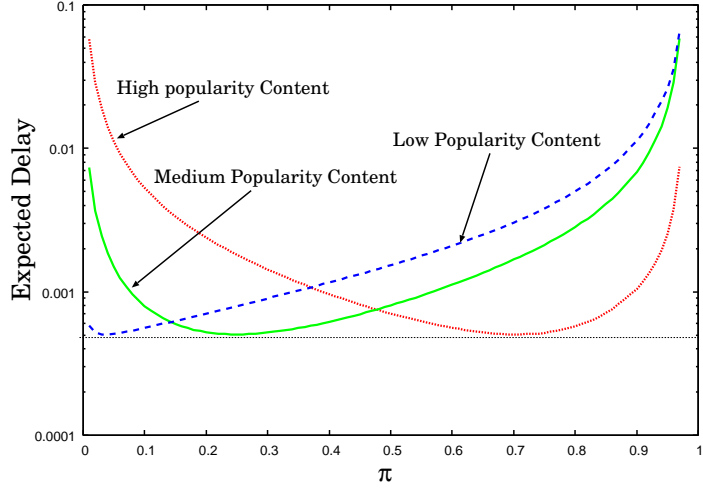} \\
\vspace{-7cm} 
(a) Minimum expected delay is obtained for $\pi_1=0.71$, $\pi_2=0.25$
and $\pi_3=0.04$  \\
\includegraphics[scale=0.5]{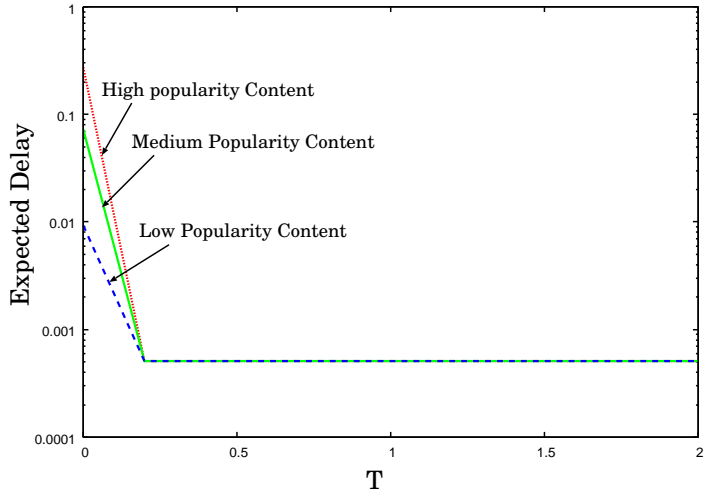} \\
\vspace{-7cm}
(b) Minimum expected delay is obtained for $T_1 \ge 0.2$, $T_2 \ge 0.2$
and $T_3 \ge 0.2$  \\
\caption{Minimum expected delay for each value of $\pi_c$ and $T_c$.}
\label{fig:3-contents}
\end{figure}

Using~\eqref{eq:d}, we compute the expected delay for different values of $\pi_c$ and $T_c$, $\pi_c$ varying from $0.01$
to $0.99$ and $T_c$ varying from $0$ to $30$s, $i=1,2,3$.   The results of our exhaustive search
 for the minimum delay are reported in 
Figure~\ref{fig:3-contents}. 
Figure~\ref{fig:3-contents}(a) shows the minimum average delay  attained as a function of 
$\pi_1$, $\pi_2$ and $\pi_3$, {\it considering all possible values of the other parameters}.
Similarly,  Figure~\ref{fig:3-contents}(b) shows
the minimum attainable average delay  as a function of $T_1$, $T_2$ and $T_3$.

For large values of  $T$, it was shown in Section~\ref{sec:optplgrout} 
 that~\eqref{squareroot} yields the optimal values of $\pi_c$.
For our experimental parameters, ~\eqref{squareroot} yields $\pi_1=0.71$, $\pi_2=0.25$ and $\pi_3=0.04$.
These values are very close to the three  points that minimize the expected delay  obtained using the exhaustive search,
 as shown in 
Figure~\ref{fig:3-contents}(a), which  
 indicates the usefulness of the closed-form expressions derived in this paper.  Even though the 
 solutions we obtained
 do not account for  joint search and placement,  they yield relevant guidelines that 
  can be effectively computed in a scalable fashion. 
 The exhaustive search for solutions took us a few hours using a Pentium IV machine, whereas the evaluation of
  the proposed closed-form expressions takes a fraction of seconds.

%% file: discussion.tex
\section{Discussion}

\label{sec:disc}

\subsection{Joint Placement and Search Optimization}

In this paper, we introduced a  new architecture, followed by a  model and its analysis that couples search
 through random walks with placement
through reinforced counters to yield simple expressions for metrics of interest.  The model allows us to pose
an optimization problem that is amenable to numerical solution.  Previous works considered heuristics
to solve the joint placement and search problem~\cite{araldo2014cost, rossini2014coupling, araldocost},
 accounting for the tradeoff between exploration and exploitation of paths towards content 
 replicas~\cite{chiocchetti2012exploit}. To the best of our knowledge, we are the first to account for such 
 a  tradeoff using random walks, which have previously been proposed in the context of peer-to-peer systems as an efficient
 way to search for content~\cite{ioannidis2008design}.  
We are also not aware of previous works that generalize the cache utility framework~\cite{dehghan2016utility, macashing}
 from a single cache to a cache network setting.

\subsection{Threats to Validity}

In this section we discuss some of the limitations and simplifying assumptions, as well as extensions subject
for future work.
%

\subsubsection{Threats to Internal Validity}
The parameters used in numerical evaluations serve to illustrate
different properties of the proposed  model. 
It remains for one to apply the proposed framework in a realistic setting,
showing how to make it scale for hundreds of contents whose popularities vary over time.
Section~\ref{sec:opt} provides 
a first step towards that goal.

\subsubsection{Threats to External Validity}
In this paper, we consider a  simple setup which allows us
to obtain an analytical model amenable to analysis. 
 The extension
to  caches with TTL replacement policy, as well as other policies such as LRU, FIFO and Random, is a subject for future
research.  

In Section~\ref{sec:opt} we focused on a single domain when analyzing the optimal placement and search problem.
  The extension 
to multiple domains under the assumption that the workload to  each domain  is Poisson is straightforward.  Nonetheless, 
validating the extent to which this assumption is valid is subject for future work.

Finally, we have focused on the placement and search strategies. We assumed throughout this paper the ZDD assumption (zero delay 
for downloads).
Accounting for the effects of   service capacities for download on system performance is out of the scope of this work.

%% file: concl.tex
\section{Conclusion}

\label{sec:concl}

Content search and placement are two of the most fundamental mechanisms that must be addressed
by any content distribution network.  In this paper, we have introduced a simple analytical 
model that couples search through random walks  and placement through a TTL-like mechanism. 
 Although the proposed model is  simple, it captures  the key tradeoffs involved
 in the choice of parameters.  Using the model, we posed an optimization problem which consists of minimizing 
 the expected delay experienced by users subject to expected storage  constraints.  The solution to 
 the optimization
 problem indicates for how long should one wait before resorting to custodians in order to download
 the desired content.   We believe that this paper is a first step towards a more foundational understanding
 of the relationship between search and placement, which is key for the efficient deployment of content centric networks.

%% file: ack.tex
\section{Acknowledgments}

\label{sec:ack}

Guilherme Domingues, E. de Souza e Silva, Rosa M. M. Leão and Daniel S. Menasché are 
partially supported by grants from CNPq and FAPERJ.  
Don Towsley is partially supported by grants from NSF.

%% file: appendix.tex
\section{Cache Insertion Rate}

\label{appcacheins}

 In this appendix we study the rate at which content is inserted into cache.  
Recall that associated with each counter and content there is a threshold $K$, such that
when the reinforced counter exceeds $K$, the corresponding content  must be stored into cache.
Next, we consider the impact of  $K$ on the cache insertion rate.
The cache insertion rate for a given content is 
%
the rate at which that content is brought into the cache.
%
Similarly, the cache eviction rate is the rate at which content is evicted from the cache. 
Due to flow balance,  in steady state the cache insertion rate equals the cache  eviction rate.

Let
{$\psi_c$ be the insertion rate}.   Recall that $\lambda_c$ and  $\mu_c$ are the request arrival rate for content $c$ and
the rate at which the counter associated to content $c$ is  decremented, respectively (Table~\ref{tab:notation}). 
{Then}
\begin{equation}
\label{eq:ins_rate}
\psi_c = \lambda_c \rho_c^K (1-\rho_c) = { \pi_c (\mu_c - \lambda_c) }
\end{equation}
%
Recall that the content miss rate is given by $\lambda_c \sum_{i=0}^K \rho_c^i (1-\rho_c) = \lambda_c (1-\pi_c)$.  
We note that, except for $K=0$, the content 
{insertion} 
rate is strictly smaller than  the content miss rate.

Let us now consider the impact of $K$ on the 
{insertion}
rate, assuming a constant miss rate.  
For a given miss rate, $\pi_c$ is determined.  Once $\pi_c$ is established, it follows from 
\eqref{eq:pic}  that  
larger values of $K$ yield smaller values of $\mu_c$.  
A decrease in $\mu_c$, in turn, causes  a reduction in the
{insertion} rate (see eq.~\eqref{eq:ins_rate}).

{
A smaller insertion rate, for the same hit ratio,
has several advantages:
(a) first, increasing the number of cache writes slows down 
servicing the requests for other contents, that is,
cache churn increases which reduces throughput ~\cite{hashcache,realitycheck, cachingworkload, martina2014unified};
(b) if flash memory is used for the cache, write operations are  
much slower than reads; 
(c) writes wear-out the flash memory; and 
(d) additional writes mean increasing power consumption.

Reducing the cache eviction rate might also lead to a reduction in network load.
 To appreciate this point, consider a scenario similar to the one presented in~\cite{dehghan2014complexity}.  
 A custodian is connected to a cache through one route, 
 and to clients through another separate 
route.  
 The link between the custodian and the cache is used only when a cache insertion is required.  The link
 between the custodian and the clients, in contrast,  is used after every cache miss, irrespectively of whether the cache miss resulted in 
 a cache insertion.      
In this case, reducing the cache insertion rate  produces a reduction in the  
 load of the link between the  custodian and the cache.  

In summary,  
larger  values of $K$ favor a reduction in   the insertion rate, which benefits system performance.  The impact of $K$ is similar
in spirit to that of $k$ in $k$-LRU~\cite{martina2014unified} and $N$ in $N$-hit caching~\cite{cachingworkload}. 

}
%
%

\section{Dual Problem For $\gamma T <<1$}

\label{appsmallgammat}

Let 
\begin{eqnarray}
K_{2,i}&=&\frac{\lambda_i}{\lambda}\mathcal{C} T \gamma \\
K_{1,i}&=&\frac{\lambda_i}{\lambda}\left(-(\mathcal{C}+T)- \mathcal{C} T \gamma \right) \\
K_{0,i}&=& (\mathcal{C}+T)
\end{eqnarray}
Let $\mathbf{1}$ be a row vector of ones. 
The  optimization problem posed in Section~\ref{sec:gammatsmall} can be stated as a   quadratic program,
\begin{eqnarray}
\min&& \frac{1}{2} \bpi^T \bQ \bpi + \bc^T \bpi \\
s.t. && \bA \bpi \leq \bb \\
&&  \mathbf{1} \bpi = B
\end{eqnarray}
where $\bQ$ is a diagonal matrix with $Q(i,i)=2 K_{2,i}$, $\bc$ is a vector with $c(i)=K_{1,i}$ and  
\begin{footnotesize}
\begin{equation}
\begin{array}{ll}
\bA = 
\underbrace{\begin{bmatrix} 
		1 & 0 & \cdots & 0 \\
		0 & \ddots & \ddots & \vdots \\
		\vdots & \ddots & \ddots & 0 \\
		0 & \cdots & 0 & 1 \\
-1 & 0 & \cdots & 0 \\
		0 & \ddots & \ddots & \vdots \\
		\vdots & \ddots & \ddots & 0 \\
		0 & \cdots & 0 & -1		
	\end{bmatrix}}_C
&
\bb=	{\begin{bmatrix} 
		1 \\
		1  \\
		\vdots\\
		1  \\
0 \\
		0 \\
		\vdots \\
		0	
	\end{bmatrix}}
\end{array}
\end{equation}
\end{footnotesize}
Note that because $\bQ$ is a positive-definite matrix, there is a unique global minimizer~\cite{boyd2004convex}.


Let $\bdelta = (\bnu, \bupsilon)$, where $\nu_i$ and $\upsilon_i$ are the Lagrange multipliers 
associated with the constraints $\pi_i \leq 1$   and the  non-negativity constraint  
 $\pi_i \ge 0$, $i=1, \ldots, C$, respectively. 
 The Lagrangian is given by
\begin{eqnarray}
\mathcal{L}(\bpi, \bdelta, \epsilon) &=& \frac{1}{2} \bpi^T \bQ \bpi + \bc^T \bpi+\bdelta^T( \bA \bpi - \bb) +\epsilon (\mathbf{1} \bpi - B) \\
&=&  \frac{1}{2} \sum_{i=1}^C  \pi_i^2 q_i + \sum_{i=1}^C  c_i \pi_i + \sum_{i=1}^C  \nu_i(\pi_i-1) + \sum_{i=1}^C  \upsilon_i(-\pi_i) + \epsilon \left(\sum_{i=1}^C  \pi_i - B\right)  \nonumber \\ \\
&=& \sum_{i=1}^C  \pi_i \left(  \frac{q_i \pi_i}{2}  + c_i + \nu_i -\upsilon_i + \epsilon \right) - \left( \sum_{i=1}^C  \nu_i \right)  - \epsilon B
\end{eqnarray}
To determine the dual function $g( \bdelta, \epsilon)$, defined as
\begin{equation}
g(\bdelta, \epsilon) = \inf_{\bpi} \mathcal{L}(\bpi, \bdelta, \epsilon)
\end{equation}
we note that 
\begin{equation} 
\nabla_{\bpi} \mathcal{L}(\bpi, \bdelta, \epsilon) = 0 \Rightarrow \bpi^{\star}  = -\bQ^{-1}(\bA^T \bdelta +\bc+ \mathbf{1}^T\epsilon)
\end{equation}
Then,
\begin{equation}
\pi^{\star}_i  = \frac{-1}{q_i} (c_i + \nu_i - \upsilon_i + \epsilon)
\end{equation}
The dual function is 
\begin{equation}
g(\bdelta, \epsilon) = -\frac{1}{2} \sum_{i=1}^C  (\pi^{\star}_i)^2{q_i} - \left(\sum_{i=1}^C  \nu_i\right) -\epsilon  B
\end{equation}

The dual problem is  also a quadratic program,
\begin{eqnarray}
\max_{\epsilon, \bdelta}&& -\frac{1}{2}  (\bpi^{\star})^T \bQ  \bpi^{\star} - \bb^T \bdelta  - B \epsilon\\
s.t. &&  \bdelta \ge 0 
\end{eqnarray}
The dual problem naturally yields an asynchronous  distributed solution~\cite{lee2015convergence}.

%% file: appendixstateful.tex
\section{Stateful Model} \label{sec:appstateful}


Two possible ways to implement  stateful searches are:
(a) when an inter-domain request arrives at a router and finds that the  request cannot be immediately satisfied,
a search is initiated and the searcher pre-selects $j$ out of the remaining $N-1$ routers
to conduct the search or;
(b) after the search is initiated, the searcher chooses the next router to visit uniformly at random,
from those that have not yet been visited before.

In this Appendix we consider the case in which routers are pre-selected at the beginning of the search.
We assume that  $\gamma$ is very large compared to the rate at which RCs are updated.

Let $J$ be  a random variable denoting the number of routers  to be visited by time $t$ excluding the first 
visited router, 
and as before, 
let $L_c$ be the number of replicas of content $c$ in the domain under consideration.
Note that as we do not allow revisits, $J \leq N-1$. 
Conditioning on $J=j$ visited routers and $L_c=l$ content replicas present in the $N-1$ possible caches to visit,
\begin{equation} 
\label{eq:R-choose}
\tilde{R}_c(t | J=j, L_c=l) = 
   (1-\pi_c)  \frac { {{N-1-l} \choose j} } {{{N-1} \choose j}}.
\end{equation}
We assume, like in Section~\ref{sec:model1}, that the search is sufficiently fast compared to the rate at which content
 is replaced.
 %
%
Replacing \eqref{eq:R-choose} into \eqref{eq:Lrepl},
\begin{eqnarray}
\label{eq:Rt-j-1}
\tilde{R}_c(t | J=j)
    &=& \sum_{l=0}^{N-1} \tilde{R}_c(t | J=j, L_c=l) { {N-1} \choose l } \pi_c^l (1-\pi_c)^{N-1-l} \label{eq:rtjeq1} \\
    &=&  \sum_{l=0}^{N-1-j}  \tilde{R}_c(t | J=j, L_c=l)  { {N-1} \choose l } \pi_c^l (1-\pi_c)^{N-1-l}  \label{eq:rtjeq2}  \\
    &=& (1-\pi_c) \sum_{l=0}^{N-1-j} { {N-1-j} \choose l } \pi_c^l (1-\pi_c)^{N-1-l}\\
    &=& (1-\pi_c)^{j+1}. \label{eq:rtjeqLast} 
\end{eqnarray}
\eqref{eq:rtjeq2} follows from \eqref{eq:rtjeq1} since $\tilde{R}_c(t | J=j)=0$ if $l>N-1-j$ 
as at least one of the $j$ routers necessarily has the content.

It is interesting to observe that \eqref{eq:rtjeqLast} and \eqref{eq:Rt-j-2} are identical,
although derived from two different sets of assumptions.

%


%% file: main.bbl
\begin{thebibliography}{10}
\expandafter\ifx\csname url\endcsname\relax
  \def\url#1{\texttt{#1}}\fi
\expandafter\ifx\csname urlprefix\endcsname\relax\def\urlprefix{URL }\fi
\expandafter\ifx\csname href\endcsname\relax
  \def\href#1#2{#2} \def\path#1{#1}\fi

\bibitem{ciscovni}
CISCO, Cisco visual networking index: Forecast and methodology (white paper),
  \url{http://www.cisco.com/c/en/us/solutions/service-provider/visual-networking-index-vni/index.html}
  (2015).

\bibitem{ccn1}
V.~Jacobson, D.~K. Smetters, J.~D. Thornton, M.~F. Plass, N.~H. Briggs, R.~L.
  Braynard, Networking named content, in: Proc. of CoNEXT, 2009, pp. 1--12.

\bibitem{dona}
T.~Koponen, M.~Chawla, B.~G. Chun, A.~Ermolinskiy, K.~H. Kim, S.~Shenker,
  I.~Stoica, A data-oriented (and beyond) network architecture, in: Proc. of
  SIGCOMM, 2007, pp. 181--192.

\bibitem{realitycheck}
D.~Perino, M.~Varvello, A reality check for content centric networking, in:
  Proceedings of the ACM SIGCOMM workshop on Information-centric networking,
  ACM, 2011, pp. 44--49.

\bibitem{incrementally2}
S.~K. Fayazbakhsh, Y.~Lin, A.~Tootoonchian, A.~Godshi, T.~Koponen, B.~M. Maggs,
  K.~Ng, V.~Sekar, S.~Schenker, Less pain, most of the gain: Incrementally
  deployable {I}{C}{N}, ACM SIGCOMM (2013) 12--16.

\bibitem{icn-comm-survey-2013}
G.~Xylomenos, C.~N. Ververidis, V.~A. Siris, N.~Fotiou, C.~Tsilopoulos,
  X.~Vasilakos, K.~V. Katsaros, G.~C. Polyzos, A survey of information-centric
  networking research, Communications Surveys and Tutorials 16~(2) (2013)
  1024--1049.

\bibitem{kurosesurvey}
J.~Kurose, Information-centric networking: The evolution from circuits to
  packets to content, Computer Networks 66 (2014) 112--120.

\bibitem{challengesicn2015}
D.~Kutscher, S.~Eum, K.~Pentikousis, I.~Psaras, D.~Corujo, D.~Saucez,
  T.~Schmidt, M.~Waehlisch, {I}{C}{N} research challenges, {I}{C}{N}{R}{G}
  ({I}{C}{N} research group - {I}{R}{T}{F}) - version 3,
  \url{https://tools.ietf.org/html/draft-irtf-icnrg-challenges-03} (2015).

\bibitem{wangpro}
L.~Wang, S.~Bayhan, J.~Ott, J.~Kangasharju, A.~Sathiaseelan, J.~Crowcroft,
  Pro-diluvian: Understanding scoped-flooding for content discovery in
  information-centric networking, in: ICN, 2015, pp. 9--18.

\bibitem{ioannidis2008design}
S.~Ioannidis, P.~Marbach, On the design of hybrid peer-to-peer systems, ACM
  SIGMETRICS Performance Evaluation Review 36~(1) (2008) 157--168.

\bibitem{psirp}
D.~Lagutin, K.~Visala, S.~Tarkoma, Publish/subscribe for internet: {PSIRP}
  perspective, in: Emerging Trends from European Research, (Valencia FIA book
  2010), 2010, pp. 75--84.

\bibitem{netinf}
Netinf, \url{http://www.4ward-project.eu/} (2010).

\bibitem{multicache}
K.~Katsaros, G.~Xylomenos, G.~C. Polyzos, Multicache: An overlay architecture
  for information-centric networking, Elsevier Computer Networks 55~(4) (2011)
  936--947.

\bibitem{elisha1}
E.~Rosensweig, J.~Kurose, D.~Towsley, Approximate models for general cache
  networks, in: Proc. of INFOCOM, 2010, pp. 1--9.

\bibitem{dabirmoghaddam2014understanding}
A.~Dabirmoghaddam, M.~M. Barijough, J.~Garcia-Luna-Aceves, Understanding
  optimal caching and opportunistic caching at the edge of information-centric
  networks, in: Proceedings of the 1st international conference on
  Information-centric networking, ACM, 2014, pp. 47--56.

\bibitem{towsley1}
N.~C. Fofack, P.~Nain, G.~Neglia, D.~Towsley, Analysis of ttl-based cache
  networks, in: Performance Evaluation Methodologies and Tools (VALUETOOLS),
  2012 6th International Conference on, IEEE, 2012, pp. 1--10.

\bibitem{gelenbe2010search}
E.~Gelenbe, Search in unknown random environments, Physical Review E 82~(6)
  (2010) 061112.

\bibitem{lv2002search}
Q.~Lv, P.~Cao, E.~Cohen, K.~Li, S.~Shenker, Search and replication in
  unstructured peer-to-peer networks, in: Proceedings of the 16th international
  conference on Supercomputing, ACM, 2002, pp. 84--95.

\bibitem{gkantsidis2004random}
C.~Gkantsidis, M.~Mihail, A.~Saberi, Random walks in peer-to-peer networks:
  algorithms and evaluation, Performance Evaluation 63~(3) (2006) 241--263.

\bibitem{traverso2013temporal}
S.~Traverso, M.~Ahmed, M.~Garetto, P.~Giaccone, E.~Leonardi, S.~Niccolini,
  Temporal locality in today's content caching: why it matters and how to model
  it, ACM SIGCOMM Computer Communication Review 43~(5) (2013) 5--12.

\bibitem{securitydht}
G.~Urdaneta, G.~Pierre, M.~V. Steen, A survey of {DHT} security techniques, ACM
  Comp. Surveys 43~(2).

\bibitem{mdht}
M.~D'Ambrosio, C.~Dannewitz, H.~Karl, V.~Vercellone, {MDHT}: A hierarchical
  name resolution service for information-centric networks, in: Proc. of
  SIGCOMM workshop on ICN, 2011, pp. 7--12.

\bibitem{mostafa}
M.~Dehghan, L.~Massoulie, D.~Towsley, D.~Menasche, Y.~Tay, A utility
  optimization approach to network cache design, in: INFOCOM, 2016.

\bibitem{mendonca2015}
G.~Mendonça, Residential nano cashe systems for video distribution (in
  portuguese), Master's thesis, COPPE/UFRJ (2015).

\bibitem{pasta-ext1992}
W.~Rosenkrantz, R.~Simba, Some theorems on conditional {PASTA}: a stochastic
  integral approach, Operations Research Letter 11 (1992) 173--177.

\bibitem{transol2000}
E.~de~Souza~e Silva, H.~R. Gail, {Transient Solutions for Markov Chains}, in:
  W.~Grassmann (Ed.), Computational Probability, Kluwer, 2000, pp. 44--79.

\bibitem{melazzi2014general}
N.~B. Melazzi, G.~Bianchi, A.~Caponi, A.~Detti, A general, tractable and
  accurate model for a cascade of lru caches, Communications Letters, IEEE
  18~(5) (2014) 877--880.

\bibitem{cohen2002replication}
E.~Cohen, S.~Shenker, Replication strategies in unstructured peer-to-peer
  networks, in: ACM SIGCOMM Computer Communication Review, Vol.~32, ACM, 2002,
  pp. 177--190.

\bibitem{liu1997static}
Z.~Liu, P.~Nain, N.~Niclausse, D.~Towsley, Static caching of web servers, in:
  Photonics West'98 Electronic Imaging, International Society for Optics and
  Photonics, 1997, pp. 179--190.

\bibitem{bay2010analytic}
L.~Bay, J.~Grau, M.~Ruiz, P.~Su, An analytic solution for some separable convex
  quadratic programming problems with equality and inequality constraints,
  Journal of Mathematical inequalities 4~(3) (2010) 453--465.

\bibitem{gallo1980quadratic}
G.~Gallo, P.~L. Hammer, B.~Simeone, Quadratic knapsack problems, in:
  Combinatorial Optimization, Springer, 1980, pp. 132--149.

\bibitem{bayon2013exact}
L.~Bay{\'o}n, J.~Grau, M.~Ruiz, P.~Su{\'a}rez, An exact algorithm for the
  continuous quadratic knapsack problem via infimal convolution, in: Handbook
  of Optimization, Springer, 2013, pp. 97--127.

\bibitem{araldo2014cost}
A.~Araldo, M.~Mangili, F.~Martignon, D.~Rossi, Cost-aware caching: optimizing
  cache provisioning and object placement in {I}{C}{N}, in: Global
  Communications Conference (GLOBECOM), IEEE, 2014, pp. 1108--1113.

\bibitem{rossini2014coupling}
G.~Rossini, D.~Rossi, Coupling caching and forwarding: Benefits, analysis, and
  implementation, in: Proceedings of the 1st international conference on
  Information-centric networking, ACM, 2014, pp. 127--136.

\bibitem{araldocost}
A.~Araldo, D.~Rossi, F.~Martignon, Cost-aware caching: Caching more (costly
  items) for less (isps operational expenditures), Parallel and Distributed
  Systems (preprint).

\bibitem{chiocchetti2012exploit}
R.~Chiocchetti, D.~Rossi, G.~Rossini, G.~Carofiglio, D.~Perino, Exploit the
  known or explore the unknown?: hamlet-like doubts in {I}{C}{N}, in:
  Proceedings of the second edition of the ICN workshop on Information-centric
  networking, ACM, 2012, pp. 7--12.

\bibitem{dehghan2016utility}
M.~Dehghan, L.~Massoulie, D.~Towsley, D.~Menasche, Y.~Tay, A utility
  optimization approach to network cache design, INFOCOM.

\bibitem{macashing}
R.~T. Ma, D.~Towsley, Cashing in on caching: On-demand contract design with
  linear pricing, CONEXT.

\bibitem{hashcache}
A.~Badam, K.~Park, V.~S. Pai, L.~L. Peterson, Hashcache: Cache storage for the
  next billion., in: NSDI, Vol.~9, 2009, pp. 123--136.

\bibitem{cachingworkload}
M.~Z. Shafiq, A.~R. Khakpour, A.~X. Liu, Characterizing caching workload of a
  large commercial content delivery network, in: INFOCOM, IEEE, 2016.

\bibitem{martina2014unified}
V.~Martina, M.~Garetto, E.~Leonardi, A unified approach to the performance
  analysis of caching systems, in: INFOCOM, 2014 Proceedings IEEE, IEEE, 2014,
  pp. 2040--2048.

\bibitem{dehghan2014complexity}
M.~Dehghan, A.~Seetharam, B.~Jiang, T.~He, T.~Salonidis, J.~Kurose, D.~Towsley,
  R.~Sitaraman, On the complexity of optimal routing and content caching in
  heterogeneous networks, arXiv preprint arXiv:1501.00216.

\bibitem{boyd2004convex}
S.~Boyd, L.~Vandenberghe, Convex optimization, Cambridge university press,
  2004.

\bibitem{lee2015convergence}
K.~Lee, R.~Bhattacharya, On the convergence analysis of asynchronous
  distributed quadratic programming via dual decomposition, arXiv preprint
  arXiv:1506.05485.

\end{thebibliography}
